\documentclass[twocolumn,pra,superscriptaddress,showpacs,aps,floatfix,noeprint]{revtex4-2}
\usepackage{graphicx,amsfonts,amssymb,amsmath}
\usepackage{textcomp}
\usepackage{esint}
\usepackage[colorlinks=true,linktocpage,bookmarks=false,citecolor=blue,linkcolor=blue,urlcolor=blue,filecolor=blue]{hyperref}
\usepackage[utf8]{inputenc}
\usepackage[T1]{fontenc}
\usepackage{lmodern}
\usepackage[capitalise]{cleveref}
\usepackage[dvipsnames]{xcolor}
\usepackage[english]{babel}
\usepackage{afterpage}

\usepackage[normalem]{ulem}

\allowdisplaybreaks

\newcommand{\Brm}{{\mathrm{B}}}

\newcommand{\Irm}{{\mathrm{I}}}

\newcommand{\kv}{{\mathbf{k}}}

\newcommand{\rv}{{\mathbf{r}}}
\newcommand{\pv}{{\mathbf{p}}}
\newcommand{\qv}{{\mathbf{q}}}

\newcommand{\Pv}{{\mathbf{P}}}

\newcommand{\betav}{{\boldsymbol {\beta}}}

\newcommand{\Nc}{\mathcal{N}}

\newcommand{\bvh}{\hat{\mathbf{b}}}
\newcommand{\bh}{\hat{b}}

\newcommand{\oh}{\hat{o}}

\newcommand{\pvh}{\hat{\mathbf{p}}}
\newcommand{\rvh}{\hat{\mathbf{r}}}

\newcommand{\Hh}{\hat{H}}

\newcommand{\Oh}{\hat{O}}

\newcommand{\nnl}{\nonumber \\}

\DeclareMathOperator{\Tr}{Tr}

\newcommand{\bra}[1]{\ensuremath{\langle#1|}}

\newcommand{\ket}[1]{\ensuremath{|#1\rangle}}

\newcommand{\braket}[1]{\ensuremath{\langle#1\rangle}}

\newcommand{\braSket}[2]{\ensuremath{\langle#1|#2\rangle}}

\graphicspath{{./figures/}}

\newcommand{\BEC}{\mathrm{BEC}}

\begin{document}

\title{Disorder in order: Localization without randomness in a cold atom system}

\author{F\'{e}lix Rose}
\email{felix.rose@m4x.org}
\affiliation{Max Planck Institute of Quantum Optics, Hans-Kopfermann-Stra\ss{e} 1, 85748 Garching, Germany} 
\affiliation{Munich Center for Quantum Science and Technology,
	Schellingstra{\ss}e 4, 80799 Munich, Germany
}
\affiliation{ Physik Department, Technische Universit{\"a}t M{\"u}nchen, James-Franck-Strasse 1, 85748 Garching, Germany} 
\author{Richard Schmidt}
\affiliation{Max Planck Institute of Quantum Optics, Hans-Kopfermann-Str. 1, 85748 Garching, Germany} 
\affiliation{Munich Center for Quantum Science and Technology (MCQST),
	Schellingstra{\ss}e 4, 80799 Munich, Germany
}
\date{January 25, 2022} 

\begin{abstract}
We present a mapping between the Edwards model of disorder describing the motion of a single particle subject to 
randomly-positioned static scatterers and the Bose polaron problem of a light quantum impurity interacting with a Bose-Einstein condensate (BEC) of heavy atoms.  The mapping offers an experimental setting to investigate the physics of Anderson localization where, by exploiting the quantum nature of the BEC, the  time evolution of the quantum impurity emulates the disorder-averaged dynamics of the Edwards model. Valid in any space dimension, the mapping can be extended to  include interacting particles,  arbitrary disorder
or confinement, and can be generalized to study  many-body localization. Moreover, the corresponding exactly-solvable disorder model offers means to benchmark variational approaches used to study polaron physics. Here, we illustrate the mapping by focusing on the case of an impurity interacting with a one-dimensional BEC  through a contact interaction.
While a simple wave function based on the expansion in the number of bath excitations misses the localization physics entirely, a  coherent state \emph{Ansatz} combined with a canonical transformation captures the 
physics of disorder and Anderson localization.

\end{abstract}
%\pacs{xxx}

\maketitle
\section{Introduction}

Coupling a particle to the collective excitations of a system with many degrees of freedom can radically alter the particle's properties. While  this paradigm has been first proposed by Landau and Pekar to describe how the interaction between electrons and lattice phonons  gives rise to quasiparticles named polarons~\cite{Landau1948a}, it has been extended to give insight into numerous systems~\cite{Duke1965a,Alexandrov1995a}, including $^3$He--$^4$He mixtures~\cite{Baym1991a}, semiconductors~\cite{Gershenson2006a} and high-temperature superconductors~\cite{Dagotto1994a}. Although described by simple models  such as the Fröhlich Hamiltonian and in spite of intensive efforts~\cite{Devreese2009a,Volosniev2015,GarciaMarch2016a,Levinsen2015a,Yoshida2018a,Shi2018b,Levinsen2021a,Dehkharghani2018a,Jager2020,Dzsotjan2020,Guenther2021a,Massignan2021a,Isaule2021a,Schmidt2021a,Franchini2021a,seetharam2020}, a comprehensive solution of the polaron problem still escapes theory. The advent of ultracold atoms allows the realization of polaron models with high tunability and brought means to probe such models, as evidenced by the recent observation of the Bose polaron spectral function in impurity-boson mixtures~\cite{Hu2016a,Jorgensen2016a,Yan2020a,PenaArdila2019a,Skou2021a}. 

In this article, we bring together the seemingly disconnected fields of polarons and disorder.
 Following Anderson's realization that quantum interference can hinder the diffusion of a particle to the point that it becomes localized~\cite{Anderson1958a}, the interplay of disorder and quantum physics has been extensively studied, revealing intricate phenomena such as magnetoresistance~\cite{Sharvin1981a,Pannetier1984a}, coherent backscattering~\cite{Kuga1984a,Wolf1985a,Albada1985a} and many-body localization \cite{Fleishman1980a,Basko2006a,Oganesyan2007,Nandkishore2015a}.
  While Anderson localization was first observed in wave systems~\cite{Dalichaouch1991a,Ye1992a}, the development of cold atom physics allowed to achieve the localization of matter waves~\cite{Billy2008a,Jendrzejewski2012a} and realize the quantum kicked rotor model~\cite{Eckardt2015a,Hainaut2019a} which can be mapped onto the Anderson model of disorder~\cite{Fishman1982a}. 

\begin{figure}[b!]
\centering
\includegraphics{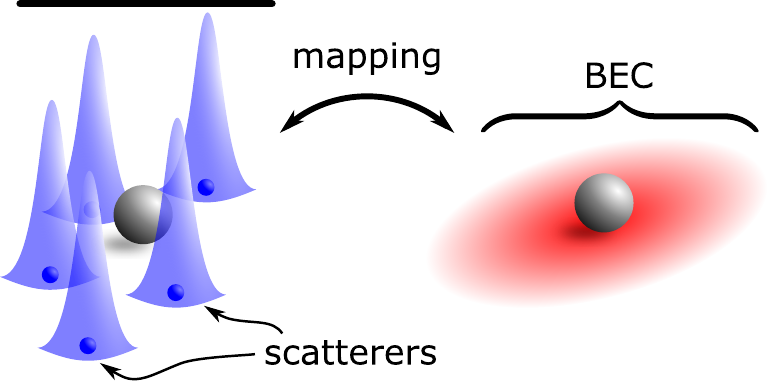}
\caption{
The disorder-averaged evolution ---represented by the overline--- of a particle 
subject to a random scattering potential (left) is mapped onto the time evolution of an impurity immersed in a homogeneous, disorder-free BEC (right).  \label{fig:mapp} }
\end{figure}

Here, we explore an alternative
way to study the physics of disorder by establishing a mapping between the disorder-averaged motion of a single particle evolving through a \emph{random disorder} potential, and that of an impurity immersed in a \emph{disorder-free} Bose-Einstein condensate (BEC). This mapping, illustrated in \cref{fig:mapp}, can experimentally be realized using a mass-imbalanced mixture of light impurities immersed in a bath of heavy bosons and provides a theoretical tool to include disorder effects in many-body approaches. 

Indeed,  both the theoretical description of disorder and Bose polarons face challenges of different origins. While the rich physics of the Bose polaron problem arises from hard to capture many-body effects, performing the disorder average for even single-particle models is challenging.  As such, a solution of the polaron problem can give insight into the corresponding disorder model, and conversely the disorder model can serve as an exactly solvable benchmark for polaron theories. 
To demonstrate this connection we show how a variational method applied to the study of  impurity models can reproduce the exact short-time solution of the corresponding disorder model and discuss implications of the mapping for studies of polaron and disorder physics.

\section{The mapping}

Many aspects of disorder physics such as Anderson localization are universal. They may depend on  dimension and symmetries 
but not on specific details of the Hamiltonian, leaving  freedom as to which model of disorder one studies. We consider the Edwards model  \cite{Edwards1958a,Akkermans2007a} that describes the evolution of a single particle through a medium of $N$ randomly-positioned static scatterers  given by the Hamiltonian
\begin{align}
\Hh_\mathrm{Ed}^{{\{\rv_i\}}} &= \dfrac{\pvh^2}{2 m_\Irm} +  V(\rvh),&
V(\rvh)&= \sum_{i=1}^N v(\rvh-\rv_i).
\label{eq:EdwModel}
\end{align}
Here $\rvh$ and $\pvh$ are the position and momentum operators of the particle and $v(\rvh-\rv_i)$ is the potential created by a scatterer at site $\rv_i$. 
In one dimension and with contact interactions, this model is known as the random Kronig--Penney model (RKPM)~\cite{Lax1958a,Nieuwenhuizen1983a}.
 The randomness comes from the positions $\rv_i$ of the scatterers, distributed e.g. uniformly in a volume $\Omega$.
 Denoting $\braket{\oh(t)}_{\{\rv_i\}}$ the expectation value of an observable $\oh$ at time $t$ for a given realization of disorder $\{\rv_i\}$, its disorder average is
\begin{equation}
\overline{\braket{\oh(t)}}=\int_{\{\rv_i\}} \braket{\oh(t)}_{\{\rv_i\}}, \label{eq:defDisAvg}
\end{equation}
where $\int_{\{\rv_i\}}=\Omega^{-N}\int d^d \rv_1 \dotsm d^d \rv_N$ is the normalized integral over all possible scatterer positions $\rv_i$. 

We now show that for any observable $\oh$ 
the disorder average~\labelcref{eq:defDisAvg} can be computed by means of a disorder-free polaron model. 
In this model one considers a system where a mobile impurity of position $\rvh$ and momentum $\pvh$ is immersed in a bosonic bath, described by the Hamiltonian 
\begin{align}\label{eq:Ham}
\Hh &=  \dfrac{\pvh^2}{2 m_\Irm}+ \sum_\kv \omega_\kv \bh^\dagger_\kv \bh_\kv +\int_{\rv'} v(\rvh-\rv') \bh^\dagger_{\rv'} \bh_{\rv'}
\end{align}
where $\omega_\kv =\kv^2/2m_\mathrm{B}$ describes the dispersion relation of bosons at momentum $\kv$ annihilated (created) by the operator  $\bh^{(\dagger)}_\kv$, and  $m_\mathrm{I}$ and $m_\mathrm{B}$ are the  masses of the impurity and the bosons, respectively. 
Both species  interact  with the density-density interaction $v(\rv)$.

To establish the mapping, we quench 
the system by preparing the impurity in a given wavepacket $\ket{\psi}$ and the bosons in a Bose-Einstein condensate (BEC) of $N$ non-interacting particles, 
\begin{equation}\label{superpos}
\ket{\BEC}=\dfrac{(\bh_{\kv=0}^\dagger)^N}{\sqrt{N!}}\ket{0} = \int_{\{\rv_i\}} \ket{\{\rv_i\}} 
\end{equation}
where
\begin{equation}
\ket{\{\rv_i\}} =  \dfrac{\bh^\dagger_{\rv_1} \dotsm \bh^\dagger_{\rv_N}}{\sqrt{N!}}\ket{0}
\end{equation}
denotes the state where the $N$ bosons have well-defined positions $\rv_i$. The combined state of the system reads
\begin{gather}
\ket{\Psi} = \ket{\psi} \otimes \ket{\BEC} = \int_{\{\rv_i\}}\ket{\psi}\otimes\ket{\{\rv_i\}}. 
\end{gather}

\begin{table}[t]
\centering
\begin{tabular}{ll}
\hline
\hline
\multicolumn{1}{c}{Polaron model} & \multicolumn{1}{c}{Disorder  model} \\
\hline
Impurity & Particle\\
Heavy boson & Static scatterer \\
Interspecies interaction & Scattering potential \\
$N$-boson state $\ket{\{\rv_i\}}$ & Disorder configuration $\{\rv_i\}$ \\
BEC state & Sampling of disorder\\
Quantum measurement $\braket{\Oh(t)}$	&	Disorder average $\overline{\braket{\oh(t)}}$ \\
\hline
\hline
\end{tabular}
\caption{Correspondence between the heavy Bose polaron and Edwards model of disorder.
\label{tab:dic}}
\end{table}

For infinitely massive bosons, $m_\Brm/m_\Irm = \infty$, the time evolution
of each state 
$\ket{\psi}\otimes\ket{\{\rv_i\}}$ contributing to the superposition 
~\labelcref{superpos} can be  determined exactly.
The boson kinetic energy drops out
and the total Hamiltonian commutes with the bosonic position operators. Hence, 
the bosons remain in the state $\ket{\{\rv_i\}}$.
Physically, the 
heavy bosons' positions are not affected by the interaction with the impurity. On the other hand, the impurity views the $N$ localized bosons as scatterers  at positions $\rv_i$ and evolves through the Edwards Hamiltonian~\labelcref{eq:EdwModel}  such that the system 
evolves into
\begin{gather}
\ket{\Psi(t)} =\int_{\{\rv_i\}} \Big[e^{-i \Hh_\mathrm{Ed}^{{\{\rv_i\}}} t}\ket{\psi}\Big]\otimes\ket{\{\rv_i\}};
\end{gather}
i.e., the system evolves as a superposition over all possible disorder realizations. 

Hence, the expectation value of any observable of the impurity $\Oh=\oh\otimes 1$ with respect to the state $\ket{\Psi(t)}$,  
\begin{equation}
\braket{\Oh(t)} = \int_{\{\rv_i\}} \bra{\psi} e^{i \Hh_\mathrm{Ed}^{{\{\rv_i\}}} t} \oh e^{-i \Hh_\mathrm{Ed}^{{\{\rv_i\}}} t} \ket{\psi} = \overline{\braket{\oh(t)}},
\end{equation}
realizes the disorder average $\overline{\braket{\oh(t)}}$~\labelcref{eq:defDisAvg}. All states in the superposition contribute equally to the measurement of $\Oh$, thus carrying out an average over all possible $\{\rv_i\}$ configurations.
 In other words, the fact that the BEC is a quantum superposition with equal weight of orthogonal states $\ket{\{\rv_i\}}$ enables disorder averaging in an analog way
 using quantum systems  such as ultracold atoms.
By contrast with a proposal in Ref.~\cite{Gavish2005a} where  disorder is simulated using pinned atoms loaded in an optical lattice, in the present work randomness is  simulated using the quantum superposition in which the bosons are prepared, similar to a suggestion made for spin systems in Ref.~\cite{Paredes2005a}.

The correspondence between the two models is summarized in \cref{tab:dic}.
We emphasize that 
$\braket{\Oh(t)}$ and $\overline{\braket{\oh(t)}}$ have very different meanings. While $\braket{\Oh(t)}$ represents a many-body measurement of the impurity evolving through the interaction with a bath of heavy bosons,  $\overline{\braket{\oh(t)}}$ corresponds to the measurement of the corresponding observable in the single-particle Edwards model averaged over many classical realizations of disorder.

\section{Generalizations} 
Most assumptions made for simplicity in the proof can be relaxed, as long as the crucial ingredient that  the $\ket{\{\rv_i\}}$ are eigenstates of the Hamiltonian remains valid.  In particular, the proof applies to any dimension.
Moreover, it is possible to include arbitrary confining potentials or interactions between the bosons, and even prepare the boson bath in a mixed state. In all these cases, the corresponding distribution of disorder would not be uniform but set by the boson state.

Crucially, the mapping holds also for 
multiple particles.  That way, it is possible to simulate 
transport properties of realistic metals
or to consider interactions between the impurities in order to investigate the interplay of many-body and disorder effects. The latter scenario has 
recently been considered in a study of
many-body localization, where the thermalization properties of mass-imbalanced mixtures was investigated~\cite{Grover2014a}. 

The mapping  holds when the bosons have a mass exceedingly large compared to the impurity. When this is not the case, the feedback of the interaction with the impurity as well as eventual interactions among bosons will induce dynamics of the bath, defining a characteristic timescale $\tau_\phi$ diverging with $m_\Brm/m_\Irm$. In the disorder model, this would correspond to dynamical disorder and cause dephasing. While this may hinder the observation of phenomena associated with disorder,
one could conversely take advantage of the presence of $\tau_\phi$ to investigate the physics of dephasing in a controlled manner. Furthermore, the paradigm of using a controlled bath as an ``auxiliary disorder'' can be reversed.
Should one be interested in the bosons' properties, 
the dynamics of the impurity can give information about the time correlations of the bosons, akin to diffusion wave spectroscopy~\cite{Berne1976a,Akkermans2007a}.

\section{Benchmarking variational solutions to  Bose polarons}

So far, we have discussed how the exact mapping can provide new approaches to disorder theory by enabling the realization of disorder in a controlled manner. However, one can also turn the correspondence around and use disorder theory to gain insight into polaron formation by providing an exactly solvable limit.  As an example, we use the mapping to test variational solutions of the Bose polaron problem. While widely used, the quality of such approximations is hard to gauge, given their nonperturbative nature as there is no small control parameter.

Specifically, we consider the one-dimensional Bose polaron described by the Hamiltonian \labelcref{eq:Ham} where a single light impurity of mass $m_\Irm$ interacts with a  bath of heavy bosons of mass $m_\Brm \gg m_\Irm$ through a  contact interaction, $v(\rv)=g\delta(\rv)$, chosen here to be repulsive ($g>0$) so that there are no boson-impurity bound states.
We turn  our attention to the spread of a particle prepared in a Gaussian wavepacket $\psi(\rv,t=0) \propto \exp(-\rv^2/2\sigma^2)$.  As discussed in \cref{AppA} the corresponding  disorder problem can be solved exactly~\cite{Nieuwenhuizen1983a}
and can thus serve as a means to assess variational approximations in the limit $m_\Brm / m_\Irm = \infty$.

\begin{figure}[t!]
\centering
\includegraphics{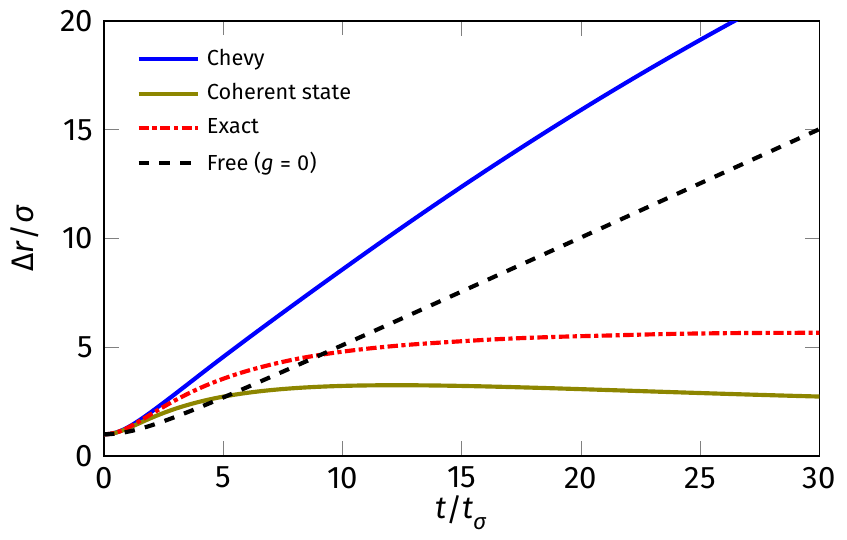}
\caption{
Time evolution of the width  $\Delta r$ of the impurity wavepacket. We compare the exact solution  (red) to  variational results based on a coherent state  (yellow) or Chevy \emph{Ansatz} (blue).  The initial width of the wavepacket $\sigma$ sets a characteristic timescale  $t_\sigma=2 m_\Irm \sigma^2/\hbar$ and the interaction and boson density $n$    are  $g/\Omega=1.5\hbar t_\sigma^{-1}$, $\sigma n=0.4$. Free spread  obtained for $g=0$ is shown for reference (black).
\label{fig:delR} }
\end{figure}

Since all states are localized in presence of disorder in one dimension, at any finite interaction $g> 0$ the wavepacket localizes at $t\to \infty$,  i.e., $\psi(\rv,t) \to  \exp(-|\rv|/2\xi)$ at large $|\rv|$ with $\xi$ a  localization length. This contrasts with the  free case ($g=0$) where 
the wavepacket spreads indefinitely.
A simple observable that distinguishes the two regimes is  the width of the wavepacket $\Delta r =\sqrt{\braket{(\rv-\braket{\rv})^2}}$. For a localized wavepacket, $\Delta r$ remains finite at all times, while in the free case $\Delta r$ grows linearly with time.

\begin{figure*}[th]
\centering
\includegraphics{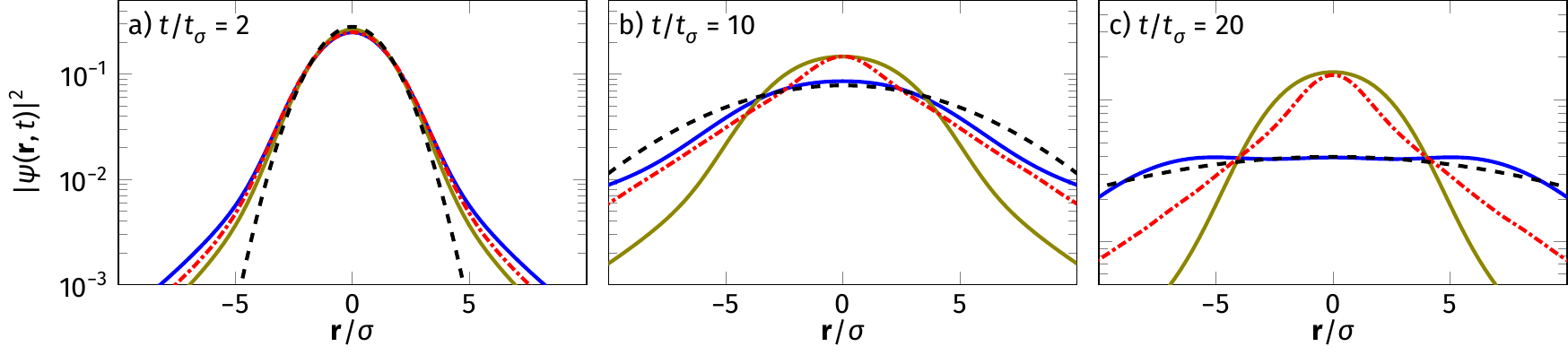}
\caption{
Density profile $|\psi(\rv,t)|^2$ as function of $\rv$ of the impurity interacting with a BEC for increasing times $t$. 
We compare the exact solution (red) 
to  the variational result based on a coherent state  (yellow) and Chevy \emph{Ansatz} (blue).  We use $\sigma n=0.4$, $g/\Omega=1.5 \hbar t_\sigma^{-1}$. The free spread ($g=0$) is shown for reference (black). 
\label{fig:compa_ex} }
\end{figure*}

We first compare the exact solution of the disorder model to the Chevy \emph{Ansatz}, which includes for the time-dependent wavefunction $\ket{\Psi^\pv(t)}$ at most one bosonic excitation \cite{Rath2013,Li2014}, i.e. 
 \begin{align}
 \ket{\Psi^\pv(t)}&{} = \alpha_0(t) \ket{\pv} \otimes \ket{\mathrm{BEC}} \nnl
 &{}+ \sum_{\qv \neq 0} \alpha_\qv(t) \ket{\pv+\qv} \otimes\bh^{\dagger}_{-\qv}\bh_0 \ket{\mathrm{BEC}}
 \label{eq:varAnsChevy}
 \end{align}
where $\alpha_\qv (t)$ are time-dependent variational parameters; for details see \cref{app:gram}. 

The  time-resolved spreading of the wavepacket is compared to the exact solution in \cref{fig:delR}.
At short times, we observe good agreement with the exact solution.
This is to be expected 
as, in variational methods, the deviation from the exact solution arises from the iterated projection of the Schr\"{o}dinger equation and the associated error did not built up  at early times. This is also reflected in the  evolution of the spatially resolved wavepacket shown in \cref{fig:compa_ex}, where the \emph{whole} density profile is well described by the Chevy  \emph{Ansatz}  at short times. Since the Chevy \emph{Ansatz} specifically relies on  including only a small number of  excitations, it is most accurate at short times at which the polaron cloud is still developing. However, the Chevy \emph{Ansatz} breaks down at larger times, and, 
rather than the localization clearly observed in the exact solution, it  predicts the indefinite spread of the wavepacket. 

This  prompts the search for more involved variational solutions.
We consider an \emph{Ansatz} based on the Lee--Low--Pines (LLP) transformation~\cite{Lee1953a} 
to transform into the comoving frame of the impurity.
 In the new frame the impurity momentum $\pv$ 
 represents the conserved total momentum of the complete system. 
 Impurity operators are hence eliminated and the bosons evolve according to the  $\pv$-dependent Hamiltonian
\begin{equation}
\Hh^\mathrm{BEC}_\pv = 
\dfrac{\big(\pv-\sum_\kv \kv \bh_\kv^\dagger \bh_\kv\big)^2}{2m_\mathrm{I}} +\sum_\kv \omega_\kv \bh^\dagger_{\kv} \bh_\kv +\dfrac{g}{\Omega}\sum_{\kv \kv'} \bh^\dagger_{\kv} \bh_{\kv'}	\label{eq:HamBECp}
\end{equation}
which now contains an transformation-induced interaction between bosons. $\Omega$ is the system volume.

The Hamiltonian \labelcref{eq:HamBECp} has been studied using various  \emph{Ansätze}~\cite{Shashi2014a,Grusdt2014a,Shchadilova2016a,Kain2017a,Nakano2017a,Yakaboylu2018a,Drescher2019a,Drescher2020a}. 
We approximate the boson wavefunction $\ket{\Phi^\pv(t)}$  by a product of coherent states,
\begin{equation}
\ket{\Phi^\pv(t)}=e^{-i \phi^\pv(t)}e^{\sum_\kv \beta^\pv_\kv(t) \bh^\dagger_\kv -\text{h.c.}}\ket{0}
\label{eq:varAnsCS}
\end{equation}
with  variational parameters $\phi^\pv(t)$ and $\beta_\kv^\pv(t)$.  By considering this wavefunction, we neglect the possibility of correlations induced between the infinitely massive bosons.

The results for the wavepacket are shown in \cref{fig:compa_ex,fig:delR}.
Again, the exact solution is well reproduced at short times,
unsurprisingly, since for a state close to the BEC, \cref{eq:varAnsCS} reduces to \cref{eq:varAnsChevy}. However, the coherent \emph{Ansatz} directly addresses one of the shortcomings of the Chevy \emph{Ansatz} by  allowing the description of a large number of bosons excited to 
small momenta and thus captures localization, although $\xi$ is underestimated.

These  examples demonstrate  how the exact mapping can help to benchmark variational solutions by providing an exact reference solution of the polaron problem. Indeed 
both the Chevy and coherent \emph{Ansätze} can be improved systematically, respectively, by allowing more bosonic excitations \cite{Levinsen2015a,Yoshida2018a,Shi2018b,Levinsen2021a} or by considering Gaussian states~\cite{Shi2018a}
with the mapping allowing to quantify the improvement. This  equally applies to other schemes such as non-Gaussian states or  quantum Monte-Carlo \cite{Ardila2015a,Ardila2016a,Grusdt2017a,Parisi2017a,Akaturk2019a,Vlietinck2015}.

\section{Application to the Anderson transition}

A long-standing question in disorder physics
 is that of the Anderson transition in three dimensions (3D). Unlike in 
 one dimension, where all states are localized, in 3D
 a mobility edge 
 separates low-energy localized
 and high-energy extended states~\cite{Anderson1958a}. 
 The  observation of the mobility edge using the expansion of atomic matter waves through disordered speckle potentials  has sparked recent theoretical~\cite{Delande2014a,Pasek2017a} and experimental interest ~\cite{Kondov2011a,Jendrzejewski2012a,Semeghini2015a}. Discrepancies between theory and experiments remain, however, as to the position of the mobility edge~\cite{Pasek2017a,Cherroret2021a}. On the theory side, the problem is difficult because it requires going beyond the perturbative weak disorder regime and treating fully the speckle potential, while for experiments it is difficult to prepare a sufficiently narrow energy distribution of
 the atomic cloud.

The polaron-disorder correspondence can help to obtain further insight. 
A specific candidate is $^6$Li impurities embedded in $^{133}$Cs with $m_\Brm/m_\Irm=22.1$
~\cite{Petrov2015a,Haefner2017a}. In particular, the Li-Cs interaction can be tuned by a Feshbach resonance at 889~G at which the Cs-Cs interaction is small ($a_{\text{CsCs}}\simeq 190a_0$). 
The  cloud of impurities expands through the bosonic medium as through a disordered Edwards potential.
Theory-wise, one does not have to deal with the speckle potential, and
can use the arsenal developed to tackle strongly interacting problems to study the  dynamics of the mixture.
Nonetheless, the question of the determination of the mobility edge remains an ambitious one. In order to answer it, two key aspects have to be addressed.
The first is given by the aforementioned decoherence induced by  the finite boson mass. 
A second constraint is given by the lifetime of the system set, e.g., by three-body loss. 
This  may  guide the choice of the system
and, for instance, molecular mixtures that feature suppressed three-body loss  might prove better candidates despite having lower mass ratios than Li-Cs \cite{ciamei2021ultracold}.

\section{Conclusion and outlook}

We have shown that a generic model of disorder, the Edwards model, is mapped to the problem of a light impurity coupled to a BEC, showing a deep connection between two seemingly  very different problems.
Our exact mapping has several implications.
Experimentally, it offers an alternative  
setup to investigate the Anderson transition. On the theory side, it can help to develop reliable
approximation schemes 
to describe the Bose polaron problem, while conversely offering an efficient way to bypass the challenging issue of disorder averaging.
However, the mapping applies to a much larger class of models. For instance, in the case of a mixture with both interactions and disorder, the mapping can be used to gauge away interactions, reducing it to a single particle problem in presence of two sources of disorder. Another possibility is the realization of more complicated models of disorder, e.g. with long-range correlations, by tuning the boson Hamiltonian. 

A promising avenue of research could originate from the field theoretical treatment of the Bose polaron problem. For instance, the Chevy \emph{Ansatz} is equivalent to a non-selfconsistent $T$-matrix approach \cite{Rath2013,Li2014}. 
It would be interesting to see how such methods compare to the diagrammatic treatment of disorder, and whether they can be used to shed light on either side of the mapping. Further intriguing applications would be the study of the interplay of disorder and interactions, especially in lower dimensions where  many-body methods such as the use of variational matrix product states could help understand thermalization and the spread of entanglement~\cite{Bardarson2012a,Serbyn2013a,Agarwal2017a,Schreiber842}. A recent proposal of a different mapping between disordered systems and interacting semi-metals brings together disorder and many-body physics in a solid state setting
~\cite{Sun2021a}.
While our work focusses on cold atom applications,  the mapping presented in this work 
can  be  extended to other condensed matter systems.
Indeed, the crucial ingredient behind the mapping is the absence of boson kinetic energy. Hence, solid state realizations using as a bath a system with flat bands ---such as encountered within the quantum Hall effect--- are also conceivable.\\

\section*{Acknowledgments}
The authors thank Cheng~Chin, Arthur~Christianen, Eugene~Demler, Falko~Pientka, Lode~Pollet, Tao~Shi, Binh~Tran,  Matteo~Zacanti and Wilhelm~Zwerger for fruitful discussions. They are grateful to 
Jean-Marc~Luck for his insight regarding the RKPM and pointing out Ref.~\cite{Nieuwenhuizen1983a}. We acknowledge support from the Deutsche Forschungsgemeinschaft (DFG, German Research Foundation) under Germany’s Excellence Strategy--EXC--2111--390814868.

\appendix

\section{Exact solution of the RKPM} \label{AppA}

The random Kronig-Penney model (RKPM) is defined by the one-dimensional Hamiltonian
\begin{equation}
\Hh_\mathrm{RKPM} = -\dfrac{\partial_x^2}{2 m} +  g\sum_{i=1}^N \delta(x-x_i),
\end{equation}
describing  a single particle in a box of size $L$ with mass $m$ and position $x\in [0,L]$ interacting with fixed random scatterers with a contact potential of strength $g$, with $\hbar$ set to $1$. We consider a given realization of disorder defined by the positions of the $N$ scatterers $x_i$, which we chose to be ennumerated as $x_{i}<x_{i+1}$ and note for convenience $x_0=0$, $x_{N+1}=L$. 

We seek to solve the Schrödinger equation
\begin{equation}
E \psi(x) = -\dfrac{1}{2 m}\psi''(x) +  g\sum_{i=1}^N \delta(x-x_i)\psi(x).
\label{eqApp:schrod}
\end{equation}
The general solution is found using a standard transfer matrix method, see e.g. Ref.~ \citep{Nieuwenhuizen1983a}. 
We consider only the case of a repulsive scattering potential, $g>0$, for which $E$ is positive, as can been seen by multiplying \cref{eqApp:schrod} by $\psi(x)^*$ and integrating over the box. On any subinterval $]x_i,x_{i+1}[$ the eigenfunctions $\psi(x)$ take the form 
\begin{equation}
\psi(x)=A_i \sin(k x + \phi_i)
\label{eqApp:psiSinWav}
\end{equation}
with $A_i$, $\phi_i$  an amplitude and a phase and $k\geq 0$  defined by $E=k^2/2m$. While $\psi$ is continuous, there is a jump in $\psi'$ at each scatterer $x_i$,  
\begin{equation}
\psi'(x_i^+)-\psi'(x_i^-)=2mg \psi(x_i)
\label{eqApp:psipJump}
\end{equation}
 as can been seen by integrating \cref{eqApp:schrod} over $[x_i^{-},x_i^{+}]$. 
 
Constructing the vector $\Psi(x)=(\psi'(x),k\psi(x))^{\mathsf{T}}$ indexed by the coordinate $x$, the conditions \labelcref{eqApp:psiSinWav,eqApp:psipJump} can be rewritten as
 \begin{align}
 \Psi(x_{i+1}^-)={}&R[k(x_{i+1}-x_i)]\Psi(x_{i}^+),&
 \Psi(x_{i}^+)={}&T\Psi(x_{i}^-),
 \end{align}
where
\begin{align}
R(\theta) = {}&\begin{pmatrix}
\cos(\theta) & -\sin(\theta) \\
\sin(\theta) & \cos(\theta) \\
\end{pmatrix}, &
T = {}&{}
\begin{pmatrix}
1 & 2mg/k \\
0 & 1 \\
\end{pmatrix}.
\end{align}
Hence, $\Psi(L) = M \Psi(0)$ with
\begin{equation}
M = R[k(L-x_N)]TR[k(x_N-x_{N-1})] \dotsm TR[k x_1].
\end{equation}
The spectrum is fixed by the boundary conditions; for instance periodic boundary conditions impose $M=1$ or equivalently $\Tr M = 2$, as $\det M=1$. 

\begin{figure*}[t]
\centering
\includegraphics{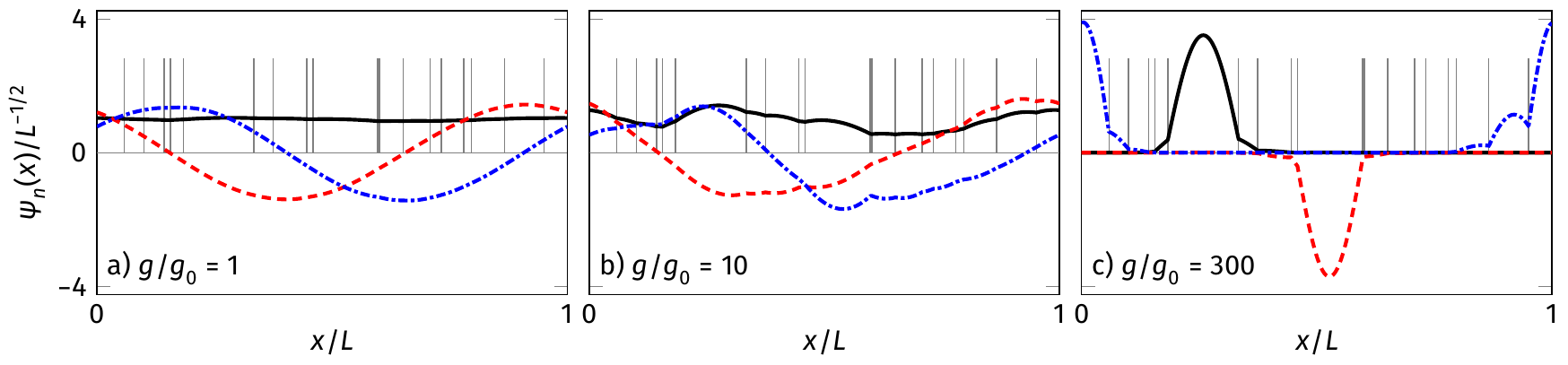}
\caption{
First three eigenstates $\psi_{n=1,2,3}(x)$ of $\Hh_\mathrm{RPKM}$ for a given realization of disorder for three different interaction strengths $g$, expressed in units of $g_0=(2mL)^{-1}$.  The ground state wavefunction and the first two excited states are shown  in solid black, dashed red and dashdotted blue, respectively. We use periodic boundary conditions in a box of size $L$. The positions of the scatterers $x_i$ are represented by grey vertical lines. For weak disorder (a) the wavefunctions closely resemble the solutions of a free particle in a box while at larger interactions (b,c) the eigenstates become localized over distances smaller than the box size.
\label{fig:EigenRPKM} }
\end{figure*}

Once the spectrum is determined, one finds for each eigenvalue the amplitudes and phases $A_i$, $\phi_i$ defining the eigenstate. Using  the continuity of $\psi(x)$ together with Eq.~\eqref{eqApp:psipJump} one relates $A_{i+1}$ and $\phi_{i+1}$ to $A_{i}$ and $\phi_{i}$. Thus by recursion 
$A_N$ and $\phi_N$ can be expressed as a function of $\phi_0$ and $A_0$. Finally, a suitable value of $\phi_0$ is determined numerically such that the boundary condition is fulfilled,
while the amplitudes are fixed by the normalization. As an illustration, we show the first eigenstates thus obtained for a given realization of disorder in \cref{fig:EigenRPKM}.

For attractive interactions ($g<0$), not considered in this article, this method remains valid. However, in this case one also needs to consider eigenstates of negative energy $E = - \kappa^2/2m$. On each subinterval, the corresponding wavefunctions then take the form $\psi(x) = A_i e^{-\kappa x} + B_i e^{\kappa x}$, physically representing states that are exponentially localized about the scatterers. The spectrum and wavefunctions can again be determined by the transfer matrix method.

\section{Time-dependent variational method}
\label{app:gram}

In this \namecref{app:gram} we briefly review how to determine the time evolution of a ket $\ket{\psi}$ under a Hamiltonian $\Hh$ using the variational method; see Refs.~[\onlinecite{McLachlan1964a,Kramer1981a,Basile1995a,Haegeman2011a,Shi2018a}] for a more in-depth discussion.
We first provide a general discussion before applying the formalism to the variational states \labelcref{eq:varAnsChevy,eq:varAnsCS}
introduced in the main text.

We seek for the best approximation of the Schr\"odinger equation $(i \partial_t-\Hh)\ket{\psi}=0$ with $\ket{\psi}$ being restrained to a variational manifold $\mathcal{M}$.
Let us assume that $\mathcal{M}=\{\ket{\psi(z_i)}, z_i \in \mathbb{C}\}$ is parameterized by  complex numbers $z_i$, with $\ket{\psi(z_i)}$ an holomorphic funtion of the $z_i$-s. Under that assumption, the following three approaches yield the same equations of motions for the variational parameters $z_i$. The first is to minimize at all times the norm of the ket $(i \partial_t-\Hh)\ket{\psi}$ with respect to the $\dot{z}_i$. The second approach is to extremize the Lagrangian
\begin{equation}
L=\dfrac{i}{2}[\bra{\psi}(\partial_t\ket{\psi})-(\partial_t\bra{\psi})\ket{\psi}]-\bra{\psi}\Hh\ket{\psi}
\end{equation}
with respect to the variational parameters $(z_i,\dot{z}_i)$. The third approach is to project at every time step the Schr\"odinger equation onto the tangent space to $\mathcal{M}$. We use in practice the last method which we present below.

\subsection{Gram matrix formulation}

 The tangent space to $\mathcal{M}$ at $\ket{\psi}$ is spanned by the vectors
\begin{align}
\ket{\partial_i \psi}=\dfrac{\partial}{\partial z_i}\ket{\psi(z)}.
\end{align}
Hence the projected Schr\"odinger equation is satisfied if and only if for all $i$
\begin{equation}
\bra{\partial_i \psi}(i \partial_t-\Hh)\ket{\psi}=0. \label{eq:projSchrod}
\end{equation}
We now define the energy of the state $E$ and the Gram matrix $G$ using the overlaps of the vectors $\ket{\partial_i \psi}$,
\begin{gather}
\begin{aligned}
E&=\bra{\psi}\Hh\ket{\psi},&
E_j&=\dfrac{\partial}{\partial z_j^*}E=\bra{\partial_j\psi}\Hh\ket{\psi}, \label{eq:gramEner}
\end{aligned}\\
G_{ij}= \braSket{\partial_i \psi}{\partial_j \psi}. \label{eq:gramMat}
\end{gather}
\cref{eq:projSchrod} can then be rewritten as
\begin{equation}
\sum_j i G_{ij} \dot{z}_j = E_i.
\end{equation}
In the specific case where $G$ is invertible the equivalent form is 
\begin{equation}
i \dot{z}_i = \sum_j [G^{-1}]_{ij}E_j. \label{eq:gramEoMinv}
\end{equation}
Otherwise there is some indeterminacy in the equations of motion, i.e. it is possible to find two different solutions $\dot{z}_i$ and $\dot{z}'_i$ fulfilling \cref{eq:gramEoMinv} provided $G_{ij}(\dot{z}_j-\dot{z}'_j)=0$.

\subsection{Application to coherent states}
\label{app:varMetCS}

While determining the equations of motion using the Gram formalism for the Chevy \emph{Ansatz} \labelcref{eq:varAnsChevy} is straightforward, the case of the coherent state \emph{Ansatz} is slightly more involved, and we present here the detailed derivation. Indeed, when applying the Gram formalism, one encounters two issues for the coherent \emph{Ansatz} \labelcref{eq:varAnsCS}. First, it is not holomorphic (as both $\beta_\kv$ and $\beta_\kv^*$ appear), and, second, there is no straight-forward equation of motion for the phase, which represents a gauge degree of freedom.

We tackle both issues at once by rather considering the variational state
\begin{align}
\ket{\beta} {}={}& \mathcal{N} e^{\sum_\kv \beta_\kv \bh^\dagger_\kv}\ket{0},
\label{eq:cohVarState}
\end{align}
parameterized by the complex numbers $\mathcal{N}$ and $\beta_\kv$, with $\mathcal{N}$ introduced for normalization. The \emph{Ansätze} \labelcref{eq:varAnsCS,eq:cohVarState} are completely equivalent provided $\Nc=e^{-i \phi - \sum_\kv |\beta_\kv|^2/2}$. 

For clarity we drop in this Section the $t$ and $\pv$ dependencies. We rewrite \cref{eq:cohVarState} by defining the vectors $(\betav)_\kv=\beta_\kv$ and $(\bvh)_\kv=\bh_\kv$, such that $\ket{\beta}=\mathcal{N} e^{\bvh^\dagger \cdot \betav}\ket{0}$. 
 We also note $S=(\betav)^\dagger \cdot \betav =  \sum_\kv |\beta_\kv|^2$.

We use the Gram matrix  formulation where from the gradients of $\ket{\beta}$,
\begin{align}
\ket{\partial_\mathcal{N} \beta}&=\dfrac{\partial}{\partial \mathcal{N}} \ket{\beta} = \dfrac{1}{\mathcal{N}}\ket{\beta},\\
\ket{\partial_\kv \beta}&=\dfrac{\partial}{\partial \beta_\kv} \ket{\beta} = \bh_\kv^\dagger\ket{\beta},
\end{align}
 we deduce the Gram matrix [\cref{eq:gramMat}]
\begin{equation}
G=
\begin{pmatrix} 
1 & \mathcal{N} \betav^\dagger  \\ 
 \mathcal{N}^*  \betav & |\mathcal{N}|^2(\delta+\betav \cdot \betav^\dagger)  
\end{pmatrix}e^{S}.
\end{equation}
In the above expression, the first line and column stand for the $\mathcal{N}$ direction, while the rest of the matrix corresponds to the momentum modes labelled by $\kv$.  In the $\kv$-$\kv$ sector, $\delta$ is understood as the identity matrix and $\betav \cdot \betav^\dagger$ as the matrix with elements $(\betav \cdot \betav^\dagger)_{\kv,\kv'}=\beta_\kv \beta_{\kv'}^*$. $G$ is invertible with the inverse 
\begin{equation}
G^{-1}=
\begin{pmatrix} 
1+S & -\dfrac{1}{\mathcal{N}^*} \betav^\dagger  \\ 
 -\dfrac{1}{\mathcal{N}}   \betav & \dfrac{1}{|\mathcal{N}|^2} \delta
\end{pmatrix}e^{-S}.
\end{equation}
The energy [\cref{eq:gramEner}] is given by
\begin{align}
E&=|\mathcal{N}|^2e^S E_0,
\end{align}
with
\begin{align}
E_0 &=  \sum_\kv (\epsilon_\kv+\omega_\kv) |\beta_\kv|^2 + \dfrac{(\pv-\Pv_\Brm[\betav])^2}{2m} + \dfrac{g}{\Omega} \Big|\sum_\qv \beta_\qv \Big|^2
\end{align}
and $\Pv_\Brm[\betav]=\sum_\kv \kv |\beta_\kv|^2$ is the total momentum of the bosons. The gradients of $E$ read
\begin{align}
E_\mathcal{N}={}& \dfrac{\partial}{\partial \mathcal{N}^*}E= \mathcal{N} e^S E_0,\\
E_\kv={}& \dfrac{\partial}{\partial \beta_\kv^*}E = |\mathcal{N}|^2 e^S 
 \bigg\{ \beta_\kv E_0  \nnl
 &{}+ \bigg[ \bigg(\epsilon_\kv+\omega_\kv-\kv \cdot \dfrac{\pv-\Pv_\Brm[\betav]}{m}\bigg)\beta_\kv+\dfrac{g}{\Omega}\sum_\qv \beta_\qv\bigg]\bigg\}.
\end{align}
Applying \cref{eq:gramEoMinv} we obtain
\begin{align}
i \dot{\beta}_\kv &=  \bigg(\epsilon_\kv+\omega_\kv-\kv \cdot \dfrac{\pv-\Pv_\Brm[\betav]}{m}\bigg)\beta_\kv+\dfrac{g}{\Omega}\sum_\qv \beta_\qv,\\
i \dot{\mathcal{N}} 		&= \mathcal{N} \dfrac{\pv^2-\Pv_\Brm[\betav]^2}{2m}.
\end{align}
From these equations of motion we deduce that $|\mathcal{N}|$ as well as $S$ are constant in time, as expected since the Hamiltonian conserves the number of particles and $S=\bra{\beta}\hat{N}\ket{\beta}/\braSket{\beta}{\beta}=N$. Hence the norm of $\ket{\beta}$, $\braSket{\beta}{\beta}=|\mathcal{N}|^2e^{S}$, is conserved. We  can rewrite $\Nc=\exp(-S/2)\exp(-i \phi)$ as
\begin{equation}
\dot\phi = \dfrac{\pv^2-\Pv_\Brm[\betav]^2}{2m_\Irm}.
\end{equation}
Using this new notation, \labelcref{eq:cohVarState} is exactly equivalent to~\labelcref{eq:varAnsCS}. The equations of motion thus obtained are the same as those present in the literature, see e.g. Ref.~\cite{Shchadilova2016a}.

\end{document}